# Accurate Stock Price Forecasting Using Robust and Optimized Deep Learning Models

Jaydip Sen
Department of Data Science
Praxis Business School
Kolkata, INDIA
email: jaydip.sen@acm.org

Sidra Mehtab
Department of Data Science
Praxis Business School
Kolkata, INDIA
email: smehtab@acm.org

*Abstract*—Designing robust frameworks for precise prediction of future prices of stocks has always been considered a very challenging research problem. The advocates of the classical *efficient market hypothesis* affirm that it is impossible to accurately predict the future prices in an efficiently operating market due to the stochastic nature of the stock price variables. However, numerous propositions exist in the literature with varying degrees of sophistication and complexity that illustrate how algorithms and models can be designed for making efficient, accurate, and robust predictions of stock prices. We present a gamut of ten deep learning models of regression for precise and robust prediction of the future prices of the stock of a critical company in the auto sector of India. Using a very granular stock price collected at 5 minutes intervals, we train the models based on the records from 31st Dec, 2012 to 27th Dec, 2013. The testing of the models is done using records from 30th Dec, 2013 to 9th Jan 2015. We explain the design principles of the models and analyze the results of their performance based on accuracy in forecasting and speed of execution.

## I. Introduction

Designing robust frameworks for precise prediction of future prices of stocks has always been considered a very challenging research problem. The advocates of the *efficient market hypothesis* affirm the impossibility of an accurate forecasting of future stock prices. However, propositions demonstrate how complex algorithms and sophisticated and optimally designed predictive models enable one to forecast future stock prices with a high degree of precision. One classical method of predicting stock prices is the *decomposition* of the stock price time series [1]. Most of the recent works in the literature for forecasting stock prices use approaches based on machine learning and deep learning [2-4]. Li et al. argue that news sentiment should be considered a vital ring in the chain of sequence from the word patterns in the news on the web to the final price movements of stocks [3]. The application of *convolutional neural networks* (CNN) in designing predictive systems for highly accurate forecasting of future stock prices is proposed in some works [4-5].

Several researchers have proposed different approaches to the technical analysis of stocks. Numerous approaches have been proposed for dealing with the technical analysis of stock prices. These approaches are based on searching and detecting well-known patterns in the sequence of stock price movement so that the investors may devise profitable strategies upon identifying appropriate patterns in the stock price data. For this purpose, a set of indicators has been recognized for characterizing the stock price patterns.

We propose a collection of forecasting models for predicting the prices of a critical stock of the automobile sector of India. The predictive framework consists of four CNN regression models and six models of regression built on the *long-and-short-term memory* (LSTM) architecture. Each model has a different architecture, different shapes of the input data, and different hyperparameter values.

The contributions of the current work are summarized as follows. First, unlike the currently existing works in the literature, which mostly deal with time-series data of daily or weekly stock prices, the models in this work are built and tested on stock price data at a small interval of 5 minutes. Second, our propositions exploit the power of deep learning, and hence, they achieve a very high degree of precision and robustness in their performance. The lowest value of the *ratio of the root mean square error* (RMSE) to the mean of the response variable is 0.00232. Finally, the speed of execution of the models is very fast. The fastest model requires 81.49 seconds for the execution of one round on the target hardware platform. It is worth mentioning here that the dataset used for training has 19500 records, while models are tested on 20500 records.

We organize the paper as follows. Section II briefly discusses some of the related work. Section III describes the data used in our work and the design details of the ten deep learning regression models. Section IV presents an extensive result of experiments on the execution and performance of the proposed models and their comparative analysis. In Section V, we conclude the paper and identify a few new directions of research.

## II. Related Work

The literature on systems and methods of stock price forecasting is quite rich. Researchers have proposed numerous techniques and approaches for accurate and robust forecasting of future movements of stock prices. Usually, these methods are organized into three types. The works of the first category use different types of regression methods, including the *ordinary least square* (OLS) regression and its other variants like *penalty-based regression*, *polynomial regression* etc. [6-8]. These approaches are not, in general, capable of handling the high degree of volatility in the stock price data. Hence, they do not lead to an acceptable level of accuracy in forecasting. The second category of propositions is based on various econometric methods like *autoregressive integrated moving average* (ARIMA), *cointegration*, *quartile regression* etc. [9-11]. The methods of this category are superior to the simple regression-based methods. However, if the stock price data is too volatile and dominated by a strong random component, the econometric methods are inadequate yielding inaccurate forecasting results. The procedures and approaches of the third category use machine learning, deep learning, and reinforcement learning algorithms, including natural language processing and sentiment analysis on the web [12-14]. These approaches are based on complex predictive models built on complex algorithms and architectures. The forecasting performances of these models are found to be most robust and accurate on volatile and granular financial data.

The most formidable challenge in designing a robust predictive model with a high level of precision for stock price forecasting is handling the randomness and the volatility exhibited by the time series. The current work utilizes the power of deep learning models in feature extraction and learning while exploiting their architectural diversity in achieving robustness and accuracy in stock price prediction on very granular time series data.

### III. METHODOLOGY

We propose a gamut of predictive models built on deep learning architectures using the historical stock prices of a company – *Hero MotoCorp*. This stock is listed in the NSE, India. The historical prices of *Hero MotoCorp* stock from 31st Dec, 2012, a Monday to 9th Jan, 2015, a Friday, are collected at 5 minutes intervals using the Metastock tool [15]. The training and validation of the models are carried out using the stock price data from 31st Dec, 2012, to 30th Dec, 2013. The testing is done using the records for the remaining period, i.e., from 31st Dec, 2013, to 9th Jan, 2015. For maintaining uniformity in the sequence, we organize the entire dataset as a sequence of daily records arranged on a weekly basis from Monday to Friday. After the data are organized, we split the dataset into a two-part – training set and the test set. While the training dataset consists of 19500 records, there are 20500 tuples in the test data. Every record has five attributes – *open*, *high*, *low*, *close*, and *volume*. We have not considered any adjusted attribute (i.e., *adjusted close*, *adjusted volume*, etc.) in our analysis.

We design ten regression models for stock price forecasting using a deep learning approach. The univariate models forecast the future values of the variable *open* based on its past values. On the other hand, the *multivariate models* predict the future values of *open* based on the historical values of all the five attributes variables in the stock data. The models are tested following an approach known as *multi-step forecasting with a walk-forward validation* [4]. In this approach, we use the training dataset for constructing the models, and the trained models are used for predicting the daily *open* values of the stock prices for the coming week. As a week completes, we include the actual stock price records of the week in the training dataset. With this extended training dataset, the *open* values are forecasted with a forecast horizon of 5 days so that the forecast for the days in the next week is available. This process continues till all the records in the test dataset are processed.

The suitability of CNNs in building predictive models for forecasting future stock prices has been demonstrated in our previous work [4]. In the current work, we present a gamut of deep learning models built on CNN and LSTM architectures and illustrate their efficacy and effectiveness in solving the same problem.

CNNs perform two critical functions for extracting rich feature sets from input data. These functions are: (1) *convolution* and (2) *pooling* or *sub-sampling* [5]. A rich set of features is extracted by the convolution operation from the input, while the sub-sampling summarizes the salient features in a given locality in the feature space. The result of the final sub-sampling in a CNN is passed on to possibly multiple dense layers. The fully-connected layers learn from the extracted features. The fully-connected layers provide the network with the power of prediction.

LSTM is an adapted form of a *recurrent neural network* (RNN) and can interpret and then forecast sequential data like text and numerical time series data [14]. The networks can maintain their state information in their designated memory cells which are also called *gates*. The state information stored in the memory cells is used in aggregating the past information available at the *forget gates*, while *input gates* receive the currently available information. The network computes the predicted value for the next time slot using it and makes it available through the output gates [14].

The ten deep learning-based models we present in this paper differ in their design, structure, and dataflows. Our proposition includes four models based on the CNN architecture and six models built on the LSTM network architecture. The proposed models are as follows. The models have been named following a convention. The first part of the model name indicates the model type (CNN or LSTM), the second part of the name indicates the nature of the input data (univariate or multivariate). Finally, the third part is an integer indicating the size of the input data to the model (5 or 10). The ten models are as follows:

(i) *CNN_UNIV_5* – a CNN model with an input of univariate *open* values of stock price records of the last week, (ii) *CNN_UNIV_10* – a CNN model with an input of univariate *open* values of stock price records of the last couple of weeks, (iii) *CNN_MULTV_10* – a CNN model with an input of multivariate stock price records consisting of five attributes of the last couple of weeks, where each variable is passed through a separate channel in a CNN, (iv) *CNN_MULTH_10* – a CNN model with the last couple of weeks' multivariate input data where each variable is used in a dedicated CNN and then combined in a *multi-headed* CNN architecture, (v) *LSTM_UNIV_5* – an LSTM with univariate *open* values of the last week as the input, (vi) *LSTM_UNIV_10* – an LSTM model with the last couple of weeks' univariate *open* values as the input, (vii) *LSTM_UNIV_ED_10* – an LSTM having an encoding and decoding ability with univariate *open* values of the last couple of weeks as the input, (viii) *LSTM_MULTV_ED_10* – an LSTM based on encoding and decoding of the multivariate stock price data of five attributes of the last couple of weeks as the input, (ix) *LSTM_UNIV_CNN_10* – a model with an encoding CNN and a decoding LSTM with univariate *open* values of the last couple of weeks as the input, and (x) *LSTM_UNIV_CONV_10* – a model having a convolutional block for encoding and an LSTM block for decoding and with univariate *open* values of the last couple of weeks as the input.

We present a brief discussion on the model design. It is important to note that all the hyperparameters (i.e., the number of nodes in a layer, the size of a convolutional, LSTM or pooling layer, etc.) used in all the models are optimized using *grid-search*. However, the parameter optimization issues are not discussed in this work.

*CNN_UNIV_5 model*: This CNN model is based on a univariate input of *open* values of the last week's stock price records. The model forecasts the following five values in the sequence as the predicted daily *open* index for the coming week. The model input has a shape (5, 1) as the five values of the last week's daily *open* index are used as the input. Since the input data for the model is too small, a solitary convolutional block and a subsequent max-subsampling block are deployed. The convolutional block has a feature space dimension of 16 and the filter (i.e., the kernel) size of 3. The convolutional block enables the model to read each input three times, and for each reading, it extracts 16 features from the input. The output data shape of the convolutional layer is (3,16). The max-pooling layer reduces the dimension of the data by a factor of 1/2. Thus, the max-pooling operation transforms the data shape to (1, 16). The result of the max-pooling layer is transformed into an array structure of one-dimension by a flattening operation. This one-dimensional vector is then passed through a dense layer block and sent to the final output layer. The output layer yields the five forecasted *open* values in sequence for the coming week. For training the model, a batch size of 4 is used with 20 epochs. The *rectified linear unit* (ReLU) *activation function* and the *Adam* optimizer for the gradient descent algorithm are used in all layers except the final output layer. In the final output layer, the *sigmoid* is used as the activation function. The use of the activation function and the optimizer is the same for all the models in this work. Fig. 1 illustrates the schematics of the model.

*CNN_UNIV_10 model*: This model is based on a univariate input of the *open* values of the last couple of weeks' stock price data. The model computes the five forecasted daily *open* values in sequence for the coming week. The structure and the dataflow for this model are identical to the *CNN_UNIV_5* model. However, the input of the model has a shape of (10,1). We use 70 epochs and 16 batch-size for training the model.

*CNN_MULTV_10 model*: This CNN model is built on the input of the last two weeks' multivariate stock price records data. The five variables of the stock price time series are used in a CNN in five separate channels. The model uses a couple of convolutional layers, each of size (32, 3). The parameter values of the convolutional blocks imply that each convolutional layer extracts features from the input data using 32 feature map size of 32 and a filter size of 3. The input to the model has a shape of (10, 5), indicating ten records, each record having five features of the stock price data. After the first convolutional operation, the shape of the data is transformed to (8, 32). The value 32 corresponds to the number of features extracted, while the value 8 is obtained by the formula: $f = (k - n) + 1$, where, $k = 10$, $n = 3$, hence, $f = 8$. Similarly, the output data shape of the second convolutional layer is (6, 32). A max-pooling layer reduces the feature space size by a factor of 1/2 producing an output data shape of (3, 32). The max-pooling block's output is then passed on to a third convolutional layer with a feature map of 16 and a kernel size of 3. The data shape of the output from the third convolutional layer becomes (1, 16) following the same computation rule. Finally, another max-pooling block receives the results of the final convolutional layer. This block does not reduce the feature space since the input data shape to it already (1, 16). Hence, and the output of the final max-pooling layer remains unchanged to (1,16). A *flatten* operation follows that converts the 16 arrays containing one value to a single array containing 16 values. The output of the *flatten* operation is passed on to a fully-connected block having 100 nodes. Finally, the output block with five nodes computes the predicted daily *open* index of the coming week. The number of epochs and the batch size used in training the model are 70 and 16, respectively. Fig. 2 depicts the *CNN_MULTV_10* model.

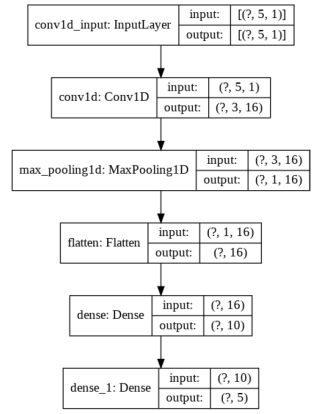

Fig. 1. The schematic structure of the model *CNN_UNIV_5*

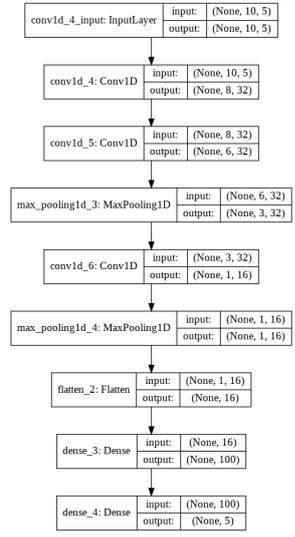

Fig. 2. The schematic structure of the model *CNN_MULTV_10*

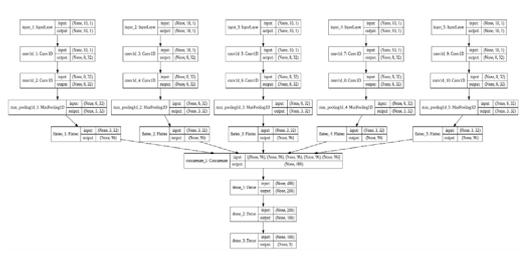

Fig. 3. The schematic structure of the model *CNN_MULTH_10*

*CNN_MULTH_10 model*: This CNN model uses a dedicated CNN block for each of the five input attributes in the stock price data. In other words, for each input variable, a separate CNN is used for feature extrication. We call this a *multivariate and multi-headed CNN model*. For each sub-CNN model, a couple of convolutional layers were used. The convolutional layers have a feature space dimension of 32 and a filter size (i.e., kernel size) of 3. The convolutional layers are followed by a max-pooling layer that reduces the feature-space size by a factor of 1/2. Following the computation rule discussed under the *CNN_MULTV_10* model, the data shape of the output from the max-pooling layer for each sub-CNN model is (3, 32). A *flatten* operation follows converting the data into a single-dimensional array of size 96 for each input variable. A *concatenation* operation follows that concatenates the five arrays, each containing 96 values, into a single one-dimensional array of size 96*5 = 480. The output of the *concatenation* operation is passed successively through two *dense layers* containing 200 nodes and 100 nodes, respectively. At the end, the output layer having five nodes yields the forecasted five values as the daily *open* stock prices for the coming week. The epoch number and the batch size used in training the model are 70 and 16, respectively. Fig. 3 shows the structure and dataflow of the *CNN_MULTH_10* model.

a dense layer) of 100 nodes. Finally, the output layer containing five nodes receives the output of the dense layer and produces the following five future values of *open* for the coming week. In training the model, 20 epochs and 16 batch-size are used. Fig. 4 presents the structure and dataflow of the model.

*LSTM_UNIV_10 model*: This univariate model uses the last couple of weeks' open index input and yields the daily forecasted *open* values for the coming week. The same values of the parameters and hyperparameters of the model *LSTM_UNIV_5* are used here. Only, the input data shape is different. The input data shape of this model is (10, 1).

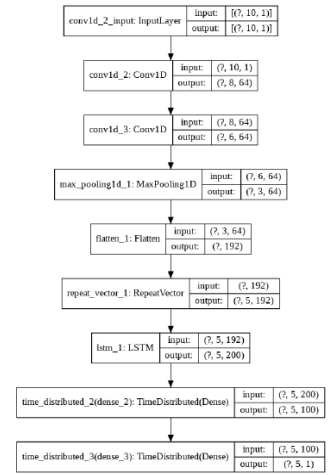

Fig. 6. The schematic structure of the model *LSTM_MULTV_CNN_10*

*LSTM_UNIV_ED_10 model*: This LSTM model has an encoding and decoding capability and is based on input of the *open* values of the stock price records of the last couple of weeks. The model consists of two LSTM blocks. One LSTM block performs the *encoding* operation, while the other does the *decoding*. The encoder LSTM block consists of 200 nodes (determined using the *grid-search* procedure). The input data shape to the encoder LSTM is (10, 1). The encoding layer yields a one-dimensional vector of size 200 – each value corresponding to the feature extracted by a node in the LSTM layer from the ten input values received from the input layer. Corresponding to each time-stamp of the output sequence (there are five time-stamps for the output sequence for the five forecasted *open* values), the input data features are extracted once. Hence, the data shape from the repeat vector layer's output is (5, 200). It signifies that 200 features are extracted from the input for each of the five time-stamps corresponding to the model's output (i.e., forecasted) sequence. The second LSTM block decodes the encoded features using 200 nodes.

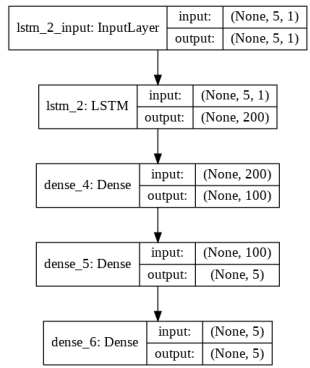

Fig. 4. The schematic structure of the model *LSTM_UNIV_5*

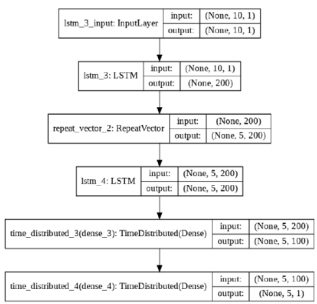

Fig. 5. The schematic structure of the model *LSTM_UNIV_ED_10*

*LSTM_UNIV_5 model*: This model is based on an input of the univariate information of the *open* values of the last week's stock price records. The model predicts the future five values in sequence as the daily *open* index for the coming week. The input has a shape of (5, 1) that indicates that the previous week's daily *open* index values are passed as the input. An LSTM block having 200 nodes receives that data from the input layer. The number of nodes at the LSTM layer is determined using the *grid-search*. The results of the LSTM block are passed on to a fully-connected layer (also known as

The decoded result is passed on to a dense layer. The dense layer learns from the decoded values and predicts the future five values of the target variable (i.e., *open*) for the coming week through five nodes in the output layer. However, the forecasted values are not produced in a single time-stamp. The forecasted values for the five days are made in five rounds. The round-wise forecasting is done using a *TimeDistributedWrapper* function that synchronizes the decoder LSTM block, the fully-connected block, and the output layer in every round. The number of epochs and the batch size used in training the model are 70 and 16, respectively. Fig. 5 presents the structure and the dataflow of the *LSTM_UNIV_ED_10* model.

*LSTM_MULTV_ED_10 model*: This model is a multivariate version of *LSTM_UNIV_ED_10*. It uses the last couple of weeks' stock price records and includes all the five attributes, i.e., *open*, *high*, *low*, *close*, and *volume*. Hence, the shape of the input data for the model is (10, 5). We use a batch size of 16 while training the model over 20 epochs.

*LSTM_UNIV_CNN_10 model*: This model is modified version of the *LSTM_UNIV_ED_N_10* model. A dedicated CNN block carries out the encoding operation. CNNs are poor in their ability to learn from sequential data. However, we exploit the power of a one-dimensional CNN in extracting important features from a time series data. After the feature extraction is done, the extracted features are provided as the input into an LSTM block. The LSTM block decodes the features and makes a robust forecasting of the future values in the sequence. The CNN block consists of a couple of convolutional layers, each of which has a feature map size of 64 and a kernel size of 3. The input data shape is (10, 1) as the model uses univariate data of the target variable of the past couple of weeks. The output shape of the initial convolutional layer is (8, 64). The value of 8 is arrived at using the computation: (10-3+1), while 64 refers to the feature space dimension.

Similarly, the shape of the output of the next convolutional block is (6, 64). A max-subsampling block follows, which contracts the feature-space dimension by 1/2. Hence, the output data shape of the max-pooling layer is (3, 64). The max-pooling layer's output is *flattened* into an array of single-dimension and size 3*64 = 192. The flattened vector is fed into the decoder LSTM block consisting of 200 nodes. The decoder architecture remains identical to the decoder block of the *LSTM_UNIV_ED_10* model. We train the model over 20 epochs with each epoch using 16 records. The structure and the dataflow of the model are shown in Fig. 6.

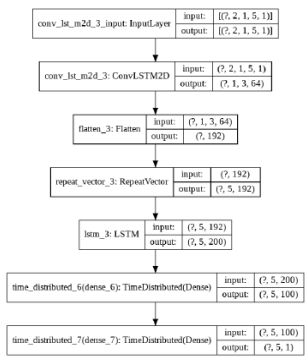

Fig. 7. The schematic structure of the model *LSTM_MULTV_CONV_10*

*LSTM_UNIV_CONV_10 model*: This model is a modification of the *LSTM_UNIV_CNN_10* model. The encoder CNN's convolution operations and the decoding operations of the LSTM sub-module are integrated for every round of the sequence in the output. This encoder-decoder model is also known as the *Convolutional-LSTM* model [13]. This integrated model reads sequential input data, performs convolution operations on the data without any explicit CNN block, and decodes the extracted features using a dedicated LSTM block. The *Keras* framework contains a class, *ConvLSTM2d*, capable of performing two-dimensional convolution operations [13]. The two-dimensional *ConvLSTM* class is tweaked to enable it to process univariate data of one-dimension. The architecture of the model *LSTM_UNIV_CONV_10* is represented in Fig. 7.

IV. EXPERIMENTAL RESULTS

This section presents an extensive set of results on the performance of the ten models. For designing a robust evaluation framework, we execute every model over ten rounds. The average performance of the ten rounds is considered as the overall performance of the model. We use four metrics for evaluation: (i) average RMSE, (ii) the RMSE for different days (i.e., Monday to Friday) of a week, (iii) the time needed for execution of one round, and (iv) the ratio of the RMSE to the response variable's (i.e., *open* value's) mean value. The models are trained on 19500 historical stock records and then tested on 20250 records. The mean value of the response variable, *open*, of the test dataset is 7386.55. All experiments are carried on a system with an Intel i7 CPU with a clock frequency in the range 2.60 GHz – 2.56 GHz and 16GB RAM. The time needed to complete one round of execution of each model is recorded in seconds. We use Python 3.7.4 on TensorFlow 2.3.0 and Keras 2.4.5 frameworks in all implementations.

TABLE I. RMSE AND EXECUTION TIME OF *CNN_UNIV_5*

| No. | Agg RMSE | Day1 | Day2 | Day3 | Day4 | Day5 | Time (sec) |
|---|---|---|---|---|---|---|---|
| 1 | 16.216 | 13.50 | 14.10 | 16.80 | 17.30 | 18.70 | 78.76 |
| 2 | 18.578 | 16.20 | 15.90 | 17.60 | 20.60 | 21.80 | 83.47 |
| 3 | 16.245 | 14.00 | 14.70 | 15.80 | 17.20 | 19.00 | 81.97 |
| 4 | 15.751 | 12.70 | 14.10 | 15.60 | 16.60 | 19.00 | 82.42 |
| 5 | 16.812 | 16.40 | 14.30 | 15.40 | 17.00 | 20.40 | 81.88 |
| 6 | 18.489 | 16.50 | 19.60 | 17.50 | 20.40 | 18.20 | 82.06 |
| 7 | 17.542 | 16.50 | 15.90 | 18.40 | 16.70 | 19.90 | 79.45 |
| 8 | 18.990 | 15.20 | 19.10 | 19.40 | 21.10 | 19.60 | 80.06 |
| 9 | 17.132 | 13.20 | 19.40 | 17.30 | 16.90 | 18.20 | 80.91 |
| 10 | 15.612 | 12.90 | 14.30 | 15.60 | 16.70 | 18.00 | 84.91 |
| **Mean** | **17.137** | 14.71 | 16.14 | 16.94 | 18.05 | 19.28 | **81.59** |
| RMSE/ Mean | **0.00232** | 0.0020 | 0.0022 | 0.0023 | 0.0024 | 0.0026 | |

Table I shows the performance of the model *CNN_UNIV_5*. The model takes, on average, 81.59 seconds to finish its one cycle. The ratio of RMSE to the average of the *open* values is 0.00232. The RMSE for day1 to day5 are 0.001991, 0.00219, 0.00229, 0.00244, and 0.00206 respectively. Here, day1 refers to Monday, and day5 is Friday. In all subsequent Tables, we will follow the same convention. In Table I, these values are listed accurately up to 4 places after the decimal due to the constraint in the column width. The RMSE values of the model *CNN_UNIV_N_5* plotted on different days in a week are depicted in Fig. 8.

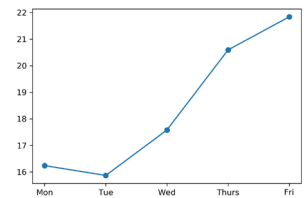

Fig. 8. RMSE vs. day plot of *CNN_UNIV_5* (depicted by the tuple#2 in Table 1)

Table II shows the results of the performance of the model *CNN_UNIV_10*. The model needs 86.99 seconds on an average for one round and produces 0.00297 as the ratio of its RMSE to the average of the actual *open* values. The daily RMSE for *day1* to *day5* are 0.00277, 0.00306, 0.00293, 0.00295, and 0.00306 respectively. Fig. 9 presents the RMSE values for the round 6 results in Table II.

TABLE II. RMSE AND EXECUTION TIME OF *CNN_UNIV_10*

| No. | Agg RMSE | Day1 | Day2 | Day3 | Day4 | Day5 | Time (sec) |
|---|---|---|---|---|---|---|---|
| 1 | 32.775 | 31.30 | 28.60 | 32.20 | 36.70 | 34.50 | 85.66 |
| 2 | 19.110 | 16.20 | 16.70 | 17.30 | 17.40 | 18.80 | 89.53 |
| 3 | 17.314 | 16.20 | 16.70 | 17.30 | 17.40 | 18.80 | 89.53 |
| 4 | 18.995 | 16.50 | 23.00 | 18.70 | 17.00 | 19.10 | 88.30 |
| 5 | 26.323 | 26.50 | 28.20 | 24.90 | 27.70 | 24.10 | 86.15 |
| 6 | 21.891 | 15.70 | 28.70 | 22.20 | 20.10 | 20.80 | 87.25 |
| 7 | 20.463 | 18.80 | 16.30 | 23.80 | 19.80 | 22.80 | 86.44 |
| 8 | 20.336 | 18.80 | 19.10 | 21.20 | 21.50 | 20.90 | 85.52 |
| 9 | 22.583 | 22.20 | 29.30 | 16.40 | 20.70 | 22.40 | 86.02 |
| 10 | 19.780 | 22.20 | 17.60 | 18.40 | 17.40 | 22.70 | 84.68 |
| **Mean** | **21.957** | 20.44 | 22.57 | 21.65 | 21.76 | 22.61 | **86.99** |
| RMSE/Mean | 0.00297 | 0.0028 | 0.0031 | 0.0029 | 0.0029 | 0.0031 | |

Table III depicts the results of *CNN_MULTV_10*. The model requires 113.49 seconds to complete its one cycle. It produces a value of 0.00380 for the ratio of the RMSE to the average of the target variable (i.e., *open* values). The RMSE values for day1 to day5 of a week are 0.00346, 0.00360, 0.00381, 0.00389, and 0.00412 respectively. The RMSE values of the model *CNN_MULTV_N_10* plotted on different days in a week are depicted in Fig. 10.

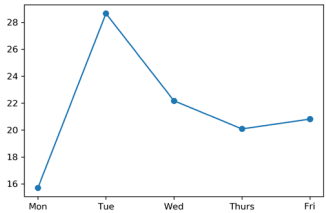

Fig. 9. The RMSE vs. day plot of *CNN_UNIV_10* (depicted by the record#6 in Table II)

TABLE III. RMSE AND EXECUTION TIME OF *CNN_MULTV_10*

| No. | Agg RMSE | Day1 | Day2 | Day3 | Day4 | Day5 | Time (sec) |
|---|---|---|---|---|---|---|---|
| 1 | 35.801 | 30.50 | 31.50 | 33.90 | 39.10 | 42.50 | 114.66 |
| 2 | 41.942 | 40.70 | 39.90 | 42.80 | 43.30 | 42.90 | 117.66 |
| 3 | 15.801 | 14.10 | 14.20 | 15.60 | 16.40 | 18.30 | 113.53 |
| 4 | 23.832 | 24.20 | 25.30 | 24.60 | 20.90 | 23.90 | 112.90 |
| 5 | 24.414 | 23.70 | 27.80 | 26.70 | 21.40 | 21.90 | 111.31 |
| 6 | 25.558 | 22.00 | 22.70 | 24.00 | 28.70 | 29.40 | 111.26 |
| 7 | 25.009 | 20.70 | 18.70 | 21.20 | 29.20 | 32.40 | 114.63 |
| 8 | 34.784 | 34.30 | 35.50 | 36.20 | 32.40 | 35.40 | 112.76 |
| 9 | 32.625 | 26.80 | 29.30 | 37.30 | 35.30 | 33.30 | 116.83 |
| 10 | 21.077 | 18.80 | 21.30 | 19.30 | 20.90 | 24.60 | 109.34 |
| **Mean** | **28.084** | 25.58 | 26.62 | 28.16 | 28.76 | 30.46 | **113.49** |
| RMSE/Mean | **0.00380** | 0.0035 | 0.0036 | 0.0038 | 0.0039 | 0.0041 | |

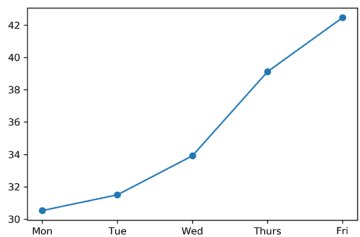

Fig. 10. RMSE vs. day plot of *CNN_MULTV_10* (depicted by the record#1 in Table III)

TABLE IV. RMSE AND EXECUTION TIME OF *CNN_MULTH_10*

| No. | Agg RMSE | Day1 | Day2 | Day3 | Day4 | Day5 | Time (sec) |
|---|---|---|---|---|---|---|---|
| 1 | 26.444 | 18.80 | 23.70 | 29.60 | 29.90 | 29.40 | 82.11 |
| 2 | 27.891 | 15.70 | 27.90 | 23.70 | 26.60 | 34.40 | 86.78 |
| 3 | 19.206 | 14.10 | 14.80 | 16.70 | 25.40 | 22.50 | 93.33 |
| 4 | 16.601 | 13.60 | 15.00 | 16.10 | 18.30 | 19.30 | 86.57 |
| 5 | 18.386 | 14.40 | 15.60 | 19.60 | 18.60 | 22.60 | 81.62 |
| 6 | 27.384 | 17.60 | 25.10 | 32.40 | 24.20 | 34.20 | 87.25 |
| 7 | 25.104 | 20.50 | 20.70 | 27.80 | 25.90 | 29.30 | 92.56 |
| 8 | 18.417 | 15.30 | 19.80 | 17.20 | 18.70 | 20.60 | 92.20 |
| 9 | 16.170 | 13.30 | 14.70 | 16.40 | 17.20 | 18.80 | 85.09 |
| 10 | 22.688 | 21.50 | 19.90 | 21.50 | 28.40 | 21.20 | 85.81 |
| **Mean** | **21.849** | 16.48 | 19.72 | 22.1 | 23.32 | 25.23 | **87.33** |
| RMSE/Mean | 0.00296 | 0.0022 | 0.0027 | 0.0030 | 0.0032 | 0.0034 | |

The results of the model *CNN_MULTH_10* are exhibited in Table IV. On average, 87.33 seconds are needed for one round of the model. The ratio of the RMSE to the average value of the target variable is 0.00296. The RMSE for day1 to day5 are, 0.00223, 0.00267, 0.00299, 0.00316, and 0.00342, respectively. The pattern of variations exhibited by the model daily RMSE is shown in Fig. 11.

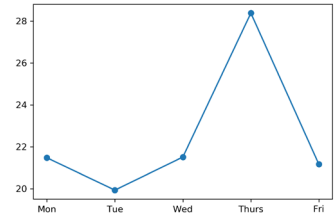

Fig. 11. The RMSE vs. day plot of *CNN_MULTH_10* (depicted by the record#10 in Table IV)

TABLE V. RMSE AND EXECUTION TIME OF *LSTM_UNIV_5*

| No. | Agg RMSE | Day1 | Day2 | Day3 | Day4 | Day5 | Time (sec) |
|---|---|---|---|---|---|---|---|
| 1 | 15.055 | 12.40 | 13.20 | 15.10 | 16.80 | 17.20 | 312.84 |
| 2 | 25.965 | 19.00 | 24.20 | 27.70 | 27.10 | 30.40 | 335.17 |
| 3 | 20.627 | 20.70 | 29.00 | 17.00 | 16.00 | 17.70 | 328.01 |
| 4 | 15.208 | 13.30 | 14.50 | 14.60 | 15.60 | 17.70 | 317.07 |
| 5 | 17.388 | 12.40 | 17.60 | 16.10 | 19.60 | 20.10 | 318.42 |
| 6 | 20.466 | 13.50 | 25.10 | 19.10 | 19.50 | 23.30 | 316.79 |
| 7 | 18.949 | 15.30 | 23.30 | 20.90 | 16.60 | 17.50 | 334.23 |
| 8 | 19.333 | 15.20 | 21.60 | 20.70 | 17.00 | 21.30 | 326.53 |
| 9 | 16.329 | 12.30 | 13.30 | 14.70 | 18.60 | 21.10 | 325.10 |
| 10 | 16.302 | 12.70 | 13.60 | 19.60 | 16.80 | 17.70 | 324.92 |
| **Mean** | **18.562.** | 14.68 | 19.54 | 18.55 | 18.36 | 20.39 | **323.17** |
| RMSE/Mean | **0.00251** | 0.0020 | 0.0026 | 0.0025 | 0.0025 | 0.0028 | |

The results of the *LSTM_UNIV_5* model are depicted in Table V. The average time to complete a round is 323.27 seconds. The ratio of the RMSE and the average value of the target variable is 0.00251. The RMSE values for day1 to day5 are 0.00199, 0.00265, 0.00251, 0.00249, and 0.00276, respectively. The pattern of variation of the daily RMSE is depicted in Fig. 12.

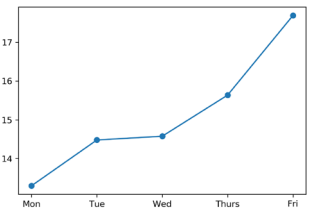

Fig. 12. The RMSE vs. day plot of *LSTM_UNIV_5* (depicted by the record#4 inf Table V)

The results of the model *LSTM_UNIV_10* are shown in Table VI. The RMSE to the average value of the target variable is 0.0092, and the model needs 532.38 seconds to complete one round. The RMSE values for day1 to day5 are 0.00243, 0.00281, 0.00289, 00305, and 0.00326 respectively. The RMSE pattern of the model is exhibited in Fig. 13.

Table VII shows that the model *LSTM_UNIV_EC_10* needs, on an average, 304.99 seconds to execute its one round. The RMSE to the average value of the target variable ratio is 0.00349. The daily RMSE values for day1 to day 5 of the model are, 0.00306, 0.00330, 0.00352, 0.00368, and 0.00382 respectively. Fig. 14 exhibits the pattern of variation of the daily RMSE.

TABLE VI. RMSE AND EXECUTION TIME OF *LSTM_UNIV_10*

| No. | Agg RMSE | Day1 | Day2 | Day3 | Day4 | Day5 | Time (sec) |
|---|---|---|---|---|---|---|---|
| 1 | 25.099 | 20.10 | 24.20 | 22.20 | 27.10 | 30.50 | 511.92 |
| 2 | 21.690 | 16.50 | 23.20 | 26.70 | 20.30 | 20.50 | 545.81 |
| 3 | 18.297 | 13.90 | 15.10 | 19.40 | 18.10 | 23.40 | 542.28 |
| 4 | 17.581 | 16.90 | 16.50 | 17.10 | 17.80 | 19.50 | 529.91 |
| 5 | 30.599 | 33.30 | 28.70 | 23.30 | 35.90 | 30.20 | 538.39 |
| 6 | 23.102 | 17.00 | 25.20 | 21.00 | 23.80 | 27.20 | 535.95 |
| 7 | 16.927 | 13.50 | 17.50 | 16.00 | 18.20 | 18.90 | 535.35 |
| 8 | 20.484 | 14.30 | 16.30 | 20.20 | 19.40 | 29.00 | 523.86 |
| 9 | 22.898 | 16.60 | 22.80 | 24.90 | 25.30 | 23.20 | 533.21 |
| 10 | 19.252 | 17.40 | 17.90 | 23.00 | 19.30 | 18.10 | 527.14 |
| Mean | **21.593** | 17.95 | 20.74 | 21.38 | 22.52 | 24.05 | **532.38** |
| RMSE/Mean | **0.00292** | 0.0024 | 0.0028 | 0.0029 | 0.0030 | 0.0033 | |

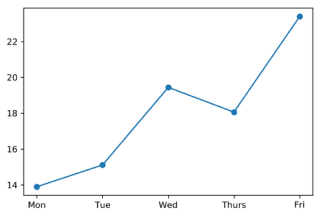

Fig. 13. The RMSE vs. day plot of *LSTM_UNIV_10* (depicted by the record#3 in Table VI)

TABLE VII. RMSE AND EXECUTION TIME OF *LSTM_UNIV_EC_10*

| No. | Agg RMSE | Day1 | Day2 | Day3 | Day4 | Day5 | Time (sec) |
|---|---|---|---|---|---|---|---|
| 1 | 31.331 | 27.30 | 30.80 | 32.70 | 32.10 | 33.30 | 302.89 |
| 2 | 29.366 | 26.40 | 28.10 | 30.00 | 30.20 | 31.80 | 307.90 |
| 3 | 38.075 | 33.10 | 36.60 | 38.00 | 41.30 | 40.80 | 305.08 |
| 4 | 24.875 | 18.80 | 22.50 | 25.00 | 27.30 | 29.40 | 307.41 |
| 5 | 22.170 | 19.60 | 20.70 | 21.50 | 23.90 | 24.70 | 305.96 |
| 6 | 19.130 | 16.80 | 17.80 | 19.30 | 20.00 | 21.40 | 306.29 |
| 7 | 21.518 | 18.70 | 19.80 | 21.40 | 22.80 | 24.40 | 302.94 |
| 8 | 19.071 | 17.10 | 17.70 | 18.90 | 19.90 | 21.50 | 308.07 |
| 9 | 30.812 | 29.20 | 30.50 | 32.20 | 30.70 | 31.30 | 303.75 |
| 10 | 21.327 | 19.20 | 19.50 | 20.80 | 23.50 | 23.30 | 299.61 |
| Mean | **25.768** | 22.62 | 24.40 | 25.98 | 27.17 | 28.19 | **304.99** |
| RMSE/Mean | **0.00349** | 0.0031 | 0.0033 | 0.0035 | 0.0037 | 0.0038 | |

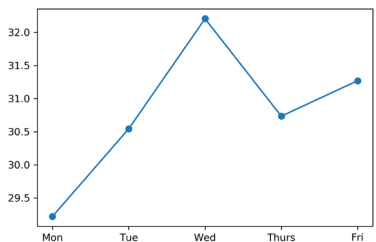

Fig. 14. The RMSE vs. day plot of *LSTM_UNIV_ED_10* (depicted by the record#9 in Table VII)

Table VIII shows that the model *LSTM_MULTV_ED_10*, on average, requires 841.29 seconds to execute one round. The ratio of the RMSE to the average value of the target is 0.00341. The daily RMSE for day1 to day5 are, respectively, 0.00316, 0.00331, 0.00341, 0.00351, and 0.00362. Fig. 15 shows the variational pattern of the daily RMSE values of the model.

TABLE VIII. RMSE AND EXECUTION TIME OF *LSTM_MULTV_EC_10*

| No. | Agg RMSE | Day1 | Day2 | Day3 | Day4 | Day5 | Time (sec) |
|---|---|---|---|---|---|---|---|
| 1 | 29.370 | 27.80 | 29.20 | 29.50 | 29.80 | 30.40 | 913.72 |
| 2 | 21.359 | 21.50 | 21.50 | 20.10 | 21.10 | 22.50 | 847.84 |
| 3 | 20.098 | 17.90 | 18.40 | 20.40 | 21.20 | 22.30 | 785.57 |
| 4 | 34.140 | 33.10 | 33.70 | 34.50 | 34.30 | 35.00 | 834.67 |
| 5 | 25.718 | 24.10 | 25.80 | 27.20 | 25.60 | 25.80 | 857.74 |
| 6 | 21.123 | 17.90 | 20.50 | 21.60 | 22.30 | 23.00 | 835.35 |
| 7 | 19.193 | 17.70 | 18.10 | 18.40 | 20.20 | 21.30 | 828.54 |
| 8 | 35.811 | 34.20 | 35.70 | 36.30 | 36.20 | 36.20 | 846.32 |
| 9 | 21.030 | 17.70 | 18.90 | 21.00 | 22.80 | 24.20 | 838.78 |
| 10 | 23.931 | 21.70 | 22.90 | 23.10 | 25.40 | 26.30 | 824.33 |
| Mean | **25.177** | 23.36 | 24.47 | 25.21 | 25.89 | 26.7 | **841.29** |
| RMSE/Mean | **0.00341** | 0.0032 | 0.0033 | 0.0034 | 0.0035 | 0.0036 | |

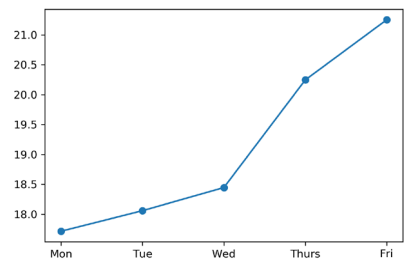

Fig. 15. The RMSE vs. day plot of *LSTM_MULTV_ED_10* (depicted by the record#7 in Table VIII)

TABLE IX. RMSE AND EXEC. TIME OF LSTM_UNIV_CNN_10

| No. | Agg RMSE | Day1 | Day2 | Day3 | Day4 | Day5 | Time (sec) |
|---|---|---|---|---|---|---|---|
| 1 | 17.415 | 14.90 | 15.90 | 17.50 | 18.50 | 19.80 | 147.18 |
| 2 | 18.929 | 16.30 | 17.50 | 19.10 | 19.70 | 21.60 | 155.53 |
| 3 | 17.300 | 14.70 | 15.90 | 17.40 | 18.30 | 19.70 | 161.90 |
| 4 | 24.030 | 21.00 | 23.50 | 25.00 | 24.60 | 25.80 | 152.65 |
| 5 | 18.859 | 16.30 | 17.20 | 18.60 | 20.20 | 21.50 | 146.21 |
| 6 | 19.380 | 17.20 | 17.50 | 19.10 | 20.30 | 22.30 | 149.56 |
| 7 | 17.609 | 15.00 | 16.10 | 17.60 | 18.50 | 20.30 | 144.24 |
| 8 | 23.994 | 22.30 | 23.50 | 23.30 | 25.30 | 25.40 | 144.67 |
| 9 | 26.790 | 24.80 | 25.70 | 27.60 | 27.30 | 28.30 | 142.94 |
| 10 | 17.730 | 15.50 | 16.30 | 17.80 | 18.60 | 20.00 | 143.08 |
| Mean | **20.204** | 17.80 | 18.91 | 20.30 | 21.13 | 22.47 | **148.80** |
| RMSE/Mean | **0.00274** | 0.0024 | 0.0026 | 0.0027 | 0.0029 | 0.0030 | |

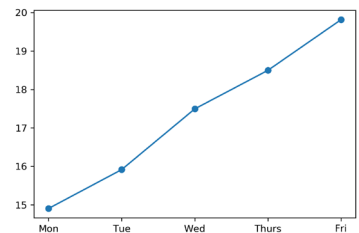

Fig. 16. The RMSE vs. day plot of *LSTM_UNIV_CNN_10* (depicted by the record#1 in Table IX)

Table IX depicts that the model *LSTM_MULTV_CNN_N_10* requires, on an average, 148.80

seconds to finish one round. The ratio of the RMSE to the average value of the target (i.e., *open*) is 0.00274. The daily RMSE values for day1 to day5 are, 0.00241, 0.00256, 0.00275, 0.00286, and 0.00304 respectively. The pattern of variation of the daily RMSE values for this model is exhibited in Fig. 16.

TABLE X.  RMSE AND EXEC. TIME OF *LSTM_UNIV_CONV_10*

| No. | Agg RMSE | Day1 | Day2 | Day3 | Day4 | Day5 | Time (sec) |
|---|---|---|---|---|---|---|---|
| 1 | 18.127 | 15.80 | 17.10 | 18.30 | 18.90 | 20.20 | 146.96 |
| 2 | 25.635 | 22.40 | 24.20 | 26.10 | 27.20 | 27.90 | 197.45 |
| 3 | 21.336 | 16.70 | 20.20 | 22.30 | 22.50 | 24.20 | 197.75 |
| 4 | 21.998 | 20.20 | 21.50 | 21.90 | 23.20 | 23.10 | 195.30 |
| 5 | 18.906 | 15.80 | 17.80 | 19.40 | 20.40 | 20.70 | 211.45 |
| 6 | 16.320 | 13.60 | 14.70 | 16.40 | 17.50 | 18.90 | 186.99 |
| 7 | 20.402 | 21.10 | 20.70 | 19.60 | 20.30 | 20.30 | 190.82 |
| 8 | 16.945 | 14.10 | 15.10 | 16.60 | 17.90 | 20.30 | 190.56 |
| 9 | 18.202 | 16.30 | 16.80 | 18.10 | 19.10 | 20.40 | 206.90 |
| 10 | 28.383 | 27.10 | 28.70 | 28.60 | 28.40 | 29.20 | 193.67 |
| **Mean** | **20.625** | 18.31 | 19.68 | 20.73 | 21.54 | 22.52 | **191.78** |
| RMSE/ Mean | **0.00279** | 0.0025 | 0.0027 | 0.0028 | 0.0029 | 0.0031 | |

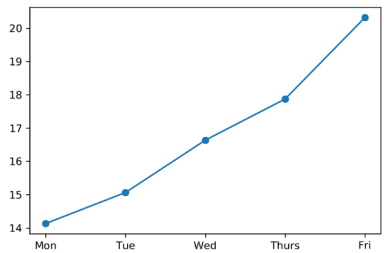

Fig. 17. The RMSE vs. day plot *LSTM_UNIV_CONV_10* (depicted by the record#8 in Table X)

The results of the model *LSTM_MULTV_CONV_N_10* are presented in Table X. The model completes its one round in 191.78 seconds. The ratio of the RMSE to the *open* value's mean is 0.00279. The daily RMSE for day1 to day5 are 0.00248, 0.00266, 0.00281, 0.00292, and 0.00305. Fig. 17 shows the patterns of daily RMSE values for this model.

TABLE XI. COMPARATIVE ANALYSIS OF THE MODELS

| RMSE/Mean | | | Time (in second) | | |
|---|---|---|---|---|---|
| Rank | Model | Value | Rank | Model | Value |
| 1 | CNN_UNIV_5 | 0.00232 | 1 | CNN_UNIV_5 | 81.59 |
| 2 | LSTM_UNIV_5 | 0.00251 | 2 | CNN_UNIV_10 | 86.99 |
| 3 | LSTM_UNIV_CNN_10 | 0.00274 | 3 | CNN_MULTH_10 | 87.33 |
| 4 | LSTM_UNIV_CONV_10 | 0.00279 | 4 | CNN_MULTV_10 | 113.49 |
| 5 | LSTM_UNIV_10 | 0.00292 | 5 | LSTM_UNIV_CNN_10 | 148.80 |
| 6 | CNN_MULTH_10 | 0.00296 | 6 | LSTM_UNIV_CNN_10 | 191.78 |
| 7 | CNN_UNIV_10 | 0.00297 | 7 | LSTM_UNIV_EC_10 | 304.99 |
| 8 | LSTM_MULTV_EC_10 | 0.00341 | 8 | LSTM_UNIV_5 | 323.27 |
| 9 | LSTM_UNIV_EC_10 | 0.00349 | 9 | LSTM_UNIV_10 | 532.38 |
| 10 | CNN_MULTV_10 | 0.00380 | 10 | LSTM_MULTV_EC_10 | 841.29 |

In Table XI, we compare the performance of the ten models we presented in this paper. We use the two metrics for ranking the models. These two metrics are: the time needed for completing one round and the ratio of the RMSE to the average of the target variable value. *CNN_UNIV_5* model is found to be the fastest and the most accurate one. In general, it is seen that the CNN models are faster in their performance, while except for two models, the LSTM models are more accurate than their CNN counterparts. These two models are *LSTM_UNIV_ED_10* and *LSTM_MULTV_ED_10*.

## V. CONCLUSION

This paper has proposed the detailed design and performance often deep learning regression models for accurate and robust prediction of daily future stock prices over a week prediction horizon. While four models are designed on CNN architecture, six models have exploited the power of LSTM architecture. The models are trained, validated, and tested on stock price records collected at 5 minutes intervals. The models are evaluated on several metrics after comprehensive rounds of experiments. The results elicited an interesting observation. It is found that while the CNN models are faster, in general, higher accuracies are yielded by the LSTM models.